\documentclass[prd, aps, preprint, floatfix, 11pt]{revtex4-1}
\usepackage{graphicx,amsmath,amsfonts,amssymb,dcolumn,epsfig,bm,float}
\usepackage{epsfig}%
\usepackage{color}
\usepackage{graphics}
\usepackage[margin=1.in]{geometry}
\usepackage[hang,small,bf]{caption}
\usepackage{rotating}
\usepackage{subfigure}
\usepackage{silence}
\def\beq{\begin{equation}}
\def\eeq{\end{equation}}
\def\bea{\begin{eqnarray}}
\def\eea{\end{eqnarray}}

\begin{document}

\begin{center} 

{\large\bf{Cosmic Acceleration in a Model of Fourth Order Gravity}}

\vskip 0.2 in

{\large{\bf Shreya Banerjee$^a$, Nilesh Jayswal$^b$ and Tejinder  P. Singh$^a$}}

\medskip

{\it $^a$Tata Institute of Fundamental Research,}
{\it Homi Bhabha Road, Mumbai 400005, India}\\
{\it $^b$Indian Institute of Technology Delhi, Hauz Khas, New Delhi 110016,  India\
}\\

\end{center}

Email: {\tt shreya.banerjee@tifr.res.in, jayswal.nilesh285@gmail.com, tpsingh@tifr.res.in}

\bigskip


\centerline{\bf ABSTRACT}
\noindent  We investigate a fourth order model of gravity, having a  free length parameter, and no cosmological constant or dark energy. We consider cosmological evolution of a flat Friedmann universe in this model for the case that the length parameter is of the order of present Hubble radius. By making a suitable choice for the present value of the Hubble parameter, and value of third derivative of the scale factor (the jerk) we find that the model can explain cosmic acceleration to the same degree of accuracy as the standard concordance model. If the free length parameter is assumed to be time-dependent, and of the order of the Hubble parameter of the corresponding epoch, the model can still explain cosmic acceleration, and provides a possible resolution of the cosmic coincidence problem. We work out the effective equation of state, and its time evolution, in our model. The fourth order correction terms are proportional to the metric, and hence mimic the cosmological constant. We also compare redshift drift in our model, with that in the standard model. The equation of state and the redshift drift serve to discriminate our model from the standard model.



\section{Introduction}
The $\Lambda$CDM model is currently the most successful theoretical explanation of cosmological observations, including CMB, cosmic acceleration, and formation and distribution of large scale structures. However, until one or more dark matter candidates are discovered in the laboratory, and/or through astronomical observations, and given also the theoretical fine tuning problems in assuming a small cosmological constant $\Lambda$, it is useful to investigate alternative explanations for non-Keplerian galaxy rotation curves and for cosmic acceleration. In particular, it is interesting to look for a common explanation for the observed galaxy rotation curves, and for cosmic acceleration, considering that the critical acceleration in the two cases is of comparable magnitude, and is of the order ${ cH_{0}}$, where ${H_0}$ is the present value of the Hubble parameter. This could be just a numerical coincidence; alternatively, it could  be an indicator of new physics.

Models of modified gravity which suggest a common origin for galaxy rotation curves and cosmic acceleration have been proposed in the literature previously. These include the TeVeS (Tensor-Vector-Scalar) theory \cite{TeVeS}, the Nu-$\Lambda$ (non-uniform cosmological constant) theory \cite{Zhao}, and its  sub-class known as the V-$\Lambda$ (Vector for $\Lambda$) model \cite{v1} \cite{v2}.

We have earlier proposed a fourth order gravity model \cite{priti1} \cite{priti2} motivated by (but independent of) the problem of averaging of Einstein's equations., and described in the next Section. The model has a free length parameter $ L$ whose value is chosen in such a way that the quantity ${ GM/L^2}$ is of the order ${ cH_0}$ where $M$ is the mass in the system under study, and ${ H_0}$ is the present value of the Hubble parameter. In the galactic case this implies that ${ L}$ is of the order of the size of the galaxy, and under these assumptions we could show that this modified gravity model implies Yukawa type corrections to the inverse square law, which leads to the non-Keplerian rotation curves  as seen in observations \cite{priti2}, without invoking dark matter. In the cosmological case, choosing ${ L=c/H_0}$ allows the universe to enter into an accelerating phase in the present epoch. A preliminary suggestion to this effect, based on an analytical solution, was made by us in \cite{priti1}. A more detailed motivation for fourth order gravity models can be found in our paper \cite{priti2}, where we also discuss that the model is not constrained by solar system tests of departures from the inverse square law. To our understanding, the issue of whether or not modified gravity models can explain cluster dynamics (especially the Bullet Cluster) is still an open one. In any case, our present discussion of cosmological considerations is independent of whether or not galactic / cluster dynamics is explained by dark matter or modified gravity.

In the present paper we examine the cosmological solution in considerable more detail than was done in \cite{priti1}. We work out the numerical solution to the modified Friedmann equations, as well as the luminosity distance - redshift relation, and compare it to Supernovae data, and show that the fit is as good as that for the $\Lambda$CDM model. The free parameter we have in hand is the third time derivative of the scale factor, which we fit to data, having in essence exchanged it for the free parameter of the $\Lambda$CDM model, namely the cosmological constant. 
We then present a more realistic version of the fourth order gravity model, where the length scale ${ L}$ is allowed to vary with epoch, and taken as ${ L=c/H}$, instead of ${L=c/H_0}$. Once again cosmic acceleration is achieved. Further, we work out the effective equation of state, as well as the redshift 
drift, in our model, and compare with the corresponding results in the standard model. These serve to discriminate between fourth order gravity and the standard model. 

Our model is of course only phenomenological, and its theoretical underpinnings remain to be discovered.   Furthermore, it is valid only in the matter dominated era as we only consider pressureless matter. In subsequent works we plan to compare theoretical predictions with other data, and also study linear perturbation theory and structure formation in fourth order gravity.

\section{Friedmann equations for Fourth Order Gravity}
In our model, Einstein's equations 
\beq
{\rm R}_{\mu\nu}-\frac{1}{2}{\rm g}_{\mu\nu}{\rm R} = \frac{8\pi {\rm G}}{{\rm c}^4}{\rm T}_{\mu\nu}
\label{1}
\eeq
are modified by adding a term containing the fourth derivative of metric tensor ${ g}_{\mu\nu}$ on the right hand side  \cite{priti1}\ \cite{priti2}
\beq
{\rm R}_{\mu\nu}-\frac{1}{2}{\rm g}_{\mu\nu}{\rm R} = {\rm \frac{8\pi G}{c^4}T}_{\mu\nu}+{\rm L}^{2}{\rm R}_{\mu\alpha\nu\beta}^{;\alpha\beta}
\label{2}
\eeq
$L$ is a length parameter which is scale dependent and defined in such a way that ${ GM/L^{2}\approx cH_{0}}$ where $M$ is the mass of the system we are working with and ${ H_{0}}$ denotes the present value of the Hubble parameter. It is not clear whether this model is derivable from an action principle. In our opinion, while it is desirable, it is not essential that the modified theory must derive from an action principle. There can be circumstances where higher order corrections can arise as effective equations (resulting say from coarse graining), in which case there may not be an underlying action principle. As pointed out by us in our earlier paper \cite{priti2} the field equations that we have considered here are motivated by (but independent of) investigations on averaging  of microscopic Einstein equations over a gravitationally polarised region. In the work of Szekeres \cite{Sz} and Zalaletdinov \cite{Za} these fourth order effective Einstein equations arise as corrections to Einstein equations owing to the existence of an underlying quadrupole moment in the mass distribution. In the present work we regard these fourth order equations as a phenomenological relic  of an underlying quantum theory of gravity, and work out their observable cosmological predictions.

Assuming a flat, homogeneous and isotropic universe on large scales, in fourth order gravity, the metric describing the dynamics of the universe, will be the spatially flat Friedmann-Lemaitre-
Robertson-Walker (FLRW) metric given by (in cartesian coordinates)
\beq
{\rm ds^{2} = c^{2}dt^{2}-a^{2}(t)(dx^{2}+dy^{2}+dz^{2})}
\label{3}
\eeq
Choosing ${ L = c/H_{0}}$ and applying the FLRW metric to the modified field equations (\ref{2}), we get the modified Friedmann equations as \cite{priti1}
\beq
\frac{\dot{{\rm a}}^{2}}{{\rm a}^{2}}+\frac{1}{{\rm H}_{0}^{2}}{\rm F}_{1}({\rm a},\dot{{\rm a}},\ddot{{\rm a}},\dddot{{\rm a}}) = {\rm \frac{8\pi G}{3}}\rho
\label{4}
\eeq
\beq
\frac{2\ddot{{\rm a}}}{{\rm a}}+\frac{\dot{{\rm a}}^{2}}{{\rm a}^{2}}+\frac{1}{{\rm H}_{0}^{2}}{\rm F}_{2}({\rm a},\dot{{\rm a}},\ddot{{\rm a}},\dddot{{\rm a}},\ddddot{{\rm a}}) = -{\rm\frac{8\pi G}{c^{2}}p}
\label{5}
\eeq
where the explicit forms of ${ F_{1}}$ and ${ F_{2}}$ are
\beq
{\rm F}_{1} = \frac{{\rm a}^{2}\dot{{\rm a}}\dddot{{\rm a}}+{\rm a}\dot{{\rm a}}^{2}\ddot{{\rm a}}-2\dot{{\rm a}}^{4}}{{\rm a}^{4}}
\label{6}
\eeq
\beq
{\rm F}_{2} = \frac{{\rm a}^{3}\ddddot{{\rm a}}+2{\rm a}^{2}\dot{{\rm a}}\dddot{{\rm a}}+{\rm a}^{2}\ddot{{\rm a}}^{2}-6{\rm a}\dot{{\rm a}}^{2}\ddot{{\rm a}}+2\dot{{\rm a}}^{4}}{{\rm a}^{4}}
\label{7}
\eeq
Writing Eqn. (\ref{5}) for a non-relativistic (i.e. pressureless) and matter-dominated universe gives
\beq
{\rm a}^{2}\dot{{\rm a}}^{2}+2{\rm a}^{3}\ddot{{\rm a}}+\frac{1}{{\rm H}_{0}^{2}}(2\dot{{\rm a}}^{4}-6{\rm a}\dot{{\rm a}}^{2}\ddot{{\rm a}}+2{\rm a}^{2}\dot{{\rm a}}\dddot{{\rm a}}+{\rm a}^{2}\ddot{{\rm a}}^{2}+{\rm a}^{3}\ddddot{{\rm a}}) = 0
\label{8}
\eeq
Since Eqn. (\ref{8}) is highly non-linear, we solve it numerically. Eqn. (\ref{4}) is to be interpreted as follows: first we use this equation to relate the present values of the first, second and third derivatives of the scale factor; a relation which is then used in Eqn. (\ref{8}). After solving for the scale factor from Eqn. (\ref{8}) one substitutes for the scale factor in Eqn. (\ref{4}) to find out the time evolution of the matter density. This requires an assumed value for the present matter density. We also note from the field equation (\ref{2}) that the Bianchi identities imply a conservation, not of the energy-momentum tensor by itself, but of the net terms on the right hand side, which include the fourth order derivative term. We will consider this issue in some detail, in Section V.

In order to simplify calculations and to make the quantities ${a, \dot{{a}}, \ddot{{ a}},\dddot{{ a}}, \ddddot{{ a}}}$ dimensionless, we introduce the following transformation of variable,
\beq 
\tau={\rm tH_{0}}
\label{8.1}
\eeq
Under this transformation, we have the following relations
\begin{center}
\bea
\dot{{\rm a}} &=& {\rm H}_{0}{\rm a}'(\tau)  \nonumber \\
 \ddot{{\rm a}} &=& {\rm H}_{0}^{2}{\rm a}''(\tau) \\
 \dddot{{\rm a}} &=& {\rm H}_{0}^{3}{\rm a}'''(\tau) \nonumber
 \label{10}
\eea
\end{center}
where prime denotes differentiation with respect to the variable $\tau$.
In terms of this new variable, Eqns. (\ref{4}) and (\ref{5}) become
\beq
\frac{{\rm a'}^{2}}{{\rm a}^{2}} + {\rm F}_{1}({\rm a},{\rm a'},{\rm a''},{\rm a'''}) = {\rm \frac{8\pi G}{3H_{0}^{2}}}\rho
\label{4new}
\eeq
\beq
\frac{2{\rm a''}}{{\rm a}} + \frac{{\rm a'}^{2}}{{\rm a}^{2}} + {\rm F}_{2}({\rm a},{\rm a'},{\rm a''},{\rm a'''},{\rm a''''}) = -{\rm\frac{8\pi G}{c^{2}H_{0}^{2}}p}
\label{5new}
\eeq
where the explicit forms of ${ F_{1}}$ and ${ F_{2}}$ are now given by
\beq
{\rm F}_{1} = \frac{{\rm a}^{2}{\rm a'}{\rm a'''} + {\rm a}{\rm a'}^{2}{\rm a''} - 2{\rm a'}^{4}}{{\rm a}^{4}}
\label{6new}
\eeq
\beq
{\rm F}_{2} = \frac{{\rm a}^{3}{\rm a''''} + 2{\rm a}^{2}{\rm a'}{\rm a'''} + {\rm a}^{2}{\rm a''}^{2} - 6{\rm a}{\rm a'}^{2}{\rm a''} + 2{\rm a'}^{4}}{{\rm a}^{4}}
\label{7new}
\eeq
The new modified Friedmann equation is given by
\beq
{\rm a^{2}a'^{2}+2a^{3}a''+2a'^{4}-6aa'^{2}a''+2a^{2}a'a'''+a^{2}a''^{2}+a^{3}a''''} = 0
\label{9}
\eeq
Eqn. (\ref{9}) is a fourth order differential equation. Hence we need three initial conditions to solve this equation. We determine the initial conditions at the present epoch ${t_{0}}$ whose numerical value we take from the Planck results \cite{planck} (which gives the present epoch of the universe using $\Lambda$CDM model). [The age of the universe in our model will be computed subsequently]. We set ${a(t_{0}) = 1}$.
${\ddot{a}}$ and ${\dddot{a}}$ are calculated by using the Taylor series expansion of the scale factor about the present epoch ${ t_{0}}$,
\begin{equation}
{\rm a(t) = a(t_{0})+\dot{a}(t_{0})[t-t_{0}]+\frac{1}{2}\ddot{a}(t_{0})[t-t_{0}]^{2}+\frac{1}{3!}\dddot{a}[t-t_{0}]^{3}+...}
\end{equation}
which can be re-expressed as
\begin{equation}
\rm{\frac{a(t)}{a(t_{0})} = 1+H_{0}[t-t_{0}]-\frac{q_{0}}{2}H_{0}^{2}[t-t_{0}]^{2}+\frac{1}{3!}j_{0}H_{0}^{3}[t-t_{0}]^{3}+...}
\end{equation}
where we have defined the deceleration parameter as
\begin{equation}
{\rm q(t) = -\frac{\ddot{a}}{a(t)H^{2}(t)}}
\label{deceleration}
\end{equation}
and the jerk parameter as
\begin{equation}
{\rm j(t) = \frac{\dddot{a}}{a(t)H^{3}(t)}}
\label{jerk}
\end{equation}
Following \cite{planck} we assume ${ t_{0}}$ = 13.798 Gyrs
and from  \cite{dec} we take ${ q_{0}} = -0.552$.  We keep the present value of the Hubble parameter (${ H_{0}}$), hence the value of ${ \dot{a}(t)}$, as a free parameter.
The present value of time (which is also the time at which the initial conditions are determined) in the  new coordinate is $\tau_{0} ={ t_{0} H_{0}}$.
Therefore the complete set of initial conditions required to solve Eqn. (\ref{9}) are given by
\begin{center}
 \bea
 {\rm a}(\tau_{0}) &=& 1 \nonumber \\
 {\rm a}'(\tau_{0}) &=& 1 \nonumber \\
 {\rm a}''(\tau_{0}) &=& 0.552 \\
 {\rm a}'''(\tau_{0}) &=& {\rm j_{0}}\nonumber
 \label{11}
 \eea
\end{center}
If we apply the above coordinate transformation to Eqn. (\ref{4}), and use the definitions of the deceleration parameter, q, and jerk parameter, j, given by Eqn. (\ref{deceleration}) and Eqn. (\ref{jerk}), respectively, we get at $\tau = \tau_{0}$ 
\begin{equation}
{\rm j_{0}-q_{0}-1} = {\rm \frac{8\pi G}{3H_{0}^{2}}}\rho =\frac{\rho}{\rho_{c}} = \Omega_{m}^{(0)}
\label{density}
\end{equation}
where $\rho_{c}$ is the critical matter density and $\Omega_{m}^{(0)}$ is the present value of matter density parameter.

Hence, once we choose a suitable value for average density $\rho$ and for ${ q_{0}}$, Eqn. (\ref{density}) implies a relation between the present values of the Hubble parameter and the jerk parameter. Again, since in the new coordinate system, the jerk parameter is given by ${ a'''}$, this in turn gives a relation between ${ a'''}$ and $H_{0}$ which can be used in the solution for Eqn. (\ref{5}) to find the best fit. Thus we have one free parameter which is the present value of the Hubble parameter.

 The present analysis has been done for two values of the average matter density of the Universe- i) for the case when we consider that the average matter density is given by the density of ordinary matter i.e. baryon density:  in this case we assume $\rho = \rho_{b} =3.347{\rm x}10^{-31}\ { \rm gm/cm^{3}}$, ii) for the case when we consider that the average matter density is given by the density of ordinary matter i.e. baryon density plus dark matter density: for this we assume a value ten times higher; $\rho = \rho_{d} =3.347{\rm x}10^{-30}\ {\rm gm/cm^{3}}$. 

In order to proceed further, we need to calculate the best fit value of the free parameter, ${ H_{0}}$. This can be done by fitting our model to the  observational data and using $\chi^{2}$ minimization method to find the value of the free parameter. In the following section, we try to find the best fit value for the free parameter.

\section{Luminosity Distance}
We find the best fit values for the model parameters by fitting the theoretical luminosity distance- redshift relation ${ d_{L}(z)}$ to the observed ${ d_{L}(z)}$ obtained from Supernovae data [Table 11 of \cite{supernova}] using the standard relation for the distance modulus $\mu$
\beq
\mu = {\rm m-M} = 5\ {\rm log}_{10}{\rm d_{L}}+25
\label{28}
\eeq
For comparison, the $\Lambda$CDM  ${ d_{L}(z)}$ relation is given by
\beq
{\rm d_{L}} = {\rm\frac{c(1+z)}{H_{0}}}\int_{0}^{{\rm z}}\frac{{\rm dz'}}{(\Omega_{m}^{(0)}(1+{\rm z'})^{3}+\Omega_{\Lambda}^{(0)})^{1/2}}
\label{25}
\eeq
where $\Omega_{m}^{(0)}$ and $\Omega_{\Lambda}^{(0)}$ are the present values of the matter density parameter and cosmological constant parameter respectively, with standard model values $\Omega_{m}^{(0)}$ = 0.32, $\Omega_{\Lambda}^{(0)}$ = 0.68, and ${ H_{0}}$ = 67.8 km/s/Mpc. In order to obtain the corresponding relation in fourth order gravity, recall that in a flat FLRW universe, the dependence of the luminosity distance on the comoving distance $\kappa$ is given by \cite{sabino} \cite{amendola} 
\beq
{\rm d_{L}} =\kappa(1+{\rm z})
\label{20.1}
\eeq
and the redshift dependence of the comoving distance can be obtained by using the null geodesic equation 
\beq
\kappa = -\int_ {{\rm t}_{0}}^{{\rm t}}\frac{{\rm c}}{{\rm a(t')}}dt'
\label{21}
\eeq
Therefore the expression for ${ d_{L}}$ now is given by
\beq
{\rm d_{L}} = {\rm\frac{c}{a(t)}}\int_{t}^{t_{0}}{\rm\frac{dt'}{a(t')}}
\label{27}
\eeq
As has already been mentioned in Section III, in order to solve Eqn. (\ref{9}), we use Eqn. (\ref{density}) for the two cases of $\rho$ mentioned there.
 Using the numerical solution for ${ a}(\tau)$ from Eqn. (\ref{9}) in Eqn. (\ref{27}), we solve it numerically in order to get ${ d_{L}}(z)$ and hence distance modulus in terms of the free parameter ${ H_{0}}$.

\begin{figure}[h!]
    \centering
        \includegraphics[width=0.75\textwidth]{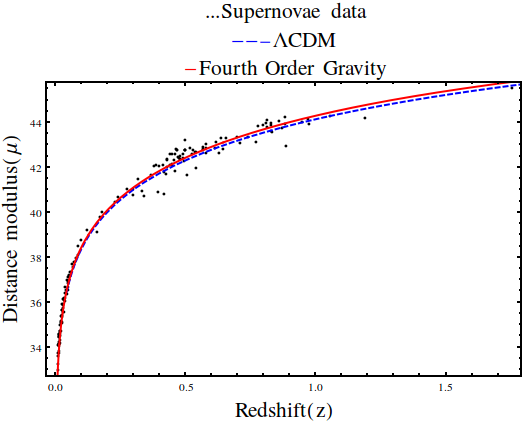}
   \caption{\small Hubble diagram for Supernovae data, $\Lambda$CDM and Fourth Order Gravity. The curve with black dots  corresponds to Supernovae data \cite{supernova}, the blue dashed curve is for $\Lambda$CDM and the red curve is for fourth order gravity.}
    \label{fig:moduluscombine}
\end{figure}
In order to find the value of the free parameter which minimizes the $\chi^{2}$ value for fourth order gravity, we calculate the reduced $\chi^{2}$ values for different values of the free parameter and for the two cases when $\rho = \rho_{b}$ and $\rho = \rho_{d}$ and find that:

\noindent for $\rho =\rho_{b}$,
\begin{equation}
\chi^{2}_{min} = 0.998 \ \ \ {\rm for}\  {\rm j_{0} = 0.4894},\ {\rm H_{0}} = 65.5^{+3}_{-1}{\rm km \ s^{-1} Mpc^{-1}}.
\label{free1}
\end{equation}
for $\rho = \rho_{d}$,
\begin{equation}
\chi^{2}_{min} = 0.998 \ \ \ {\rm for}\  {\rm j_{0} = 0.8693},\ {\rm H_{0}} = 65.0^{+3}_{-1}{\rm km \ s^{-1} Mpc^{-1}}.
\label{free2}
\end{equation}
In our further calculations we will neglect the uncertainty and take the best fit value of the Hubble parameter as ${ H_{0}} = 65.5\ { \rm km \ s^{-1} Mpc^{-1}}$ and ${ H_{0}} = 65.0\ {\rm km \ s^{-1} Mpc^{-1}}$ for the two cases respectively.
We have also calculated the $\chi^{2}$ value for $\Lambda$CDM model, which comes out to be 
\begin{equation}
\chi^{2} = 0.998.
\end{equation}
Comparing the $\chi^{2}$ values for the two models, we can say that the fourth order gravity model fits the Supernovae data as  good as the $\Lambda$CDM model.

The comparison of the models with observation is shown in  Fig. \ref{fig:moduluscombine} where the distance modulus graph for the fourth order gravity model has been plotted for the case when $\rho = \rho_{b}$. 

Supernovae observations put some constraint on the value of the third derivative of the scale factor, defined as the jerk parameter, at the present epoch. But since it is very difficult to measure the jerk, being related to the third term in the Taylor series expansion of the scale factor, direct observational
constraints are relatively very weak. The dimensionless quantity jerk is generally defined as ${  j = \dddot{a}/aH^{3}}$. In \cite{riess} \cite{visser}, the authors have reported that the jerk ${ j_{0}}$ at the present epoch, is positive at $95 \%$ confidence level. The allowed region of the jerk value (as quoted in \cite{visser}) is around (-$0.1,+6.4)$. 
A slightly different analysis (using the same raw data analysed in somewhat different
fashion) is presented by \cite{cadwell}, where the allowed range of the jerk is (-$0.5,+3.9)$.

In our model, the jerk parameter, in our new coordinate system i.e. $\tau = { tH_{0}}$, is given by ${ a'''}$. Hence, if we take into consideration the  values of  ${ a'''}  = 0.4894\ {\rm (for\ \rho_{b})}$ and  ${ a'''}  = 0.86930\ {\rm (for\ \rho_{d})}$, then these values lie well within the allowed parameter space for the jerk value from the Supernovae observations. It is also consistent with the more recent analysis of the jerk value done by \cite{zhai}. For reference, in $\Lambda$CDM model, ${ j_{0}}= 1$.

\begin{Large}
\section{Numerical exact solution of the modified Friedmann equations}
\end{Large}

\subsection{Results and Discussions}
Using the best fit free parameter values as calculated in Section III, we solve Eqn. (\ref{9})  numerically.
\begin{figure}[ht]
\centering
\includegraphics[scale=0.75]{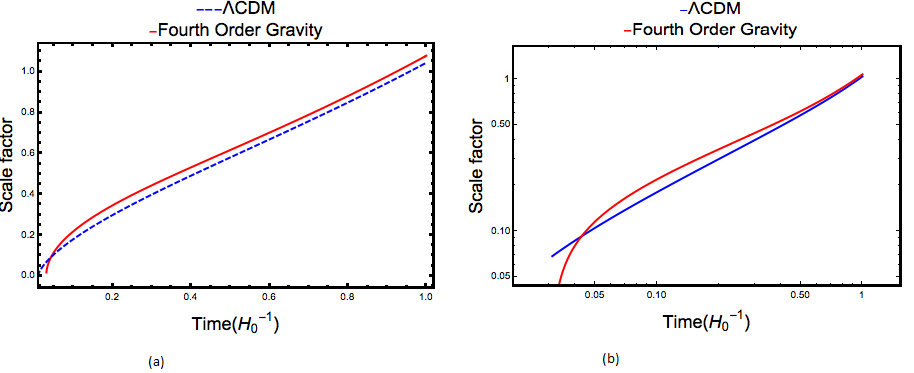}
\caption{\small a) Variation of scale factor (a(t)) with time (from $0.0\ { H_{0}^{-1}}$ to $1.0\ { H_{0}^{-1}}$) for Fourth Order Gravity and $\Lambda$CDM model. The red curve is for Fourth Order Gravity and the blue dashed curve is for $\Lambda$CDM model. b) Log-Log plot of the variation of scale factor with time (from $0.0\ { H_{0}^{-1}}$ to $1.0\ { H_{0}^{-1}}$) for Fourth Order Gravity and $\Lambda$CDM model. The red curve is for Fourth Order Gravity and the blue curve is for $\Lambda$CDM model.   
}
\label{fig:fogscalefact}
\end{figure}
Fig. \ref{fig:fogscalefact} shows the plot of the variation of scale factor with time for fourth order gravity for $\rho = \rho_{b}$. From Fig. \ref{fig:fogscalefact}, we can see that the scale factor becomes nearly zero when $\tau = 0$. Thus using the relation $\tau_{0} = { t_{0} H_{0}}$, with $\tau_{0} = 0.92348$ and ${ H_{0}} = 65.5\ {\rm km \ s^{-1}Mpc^{-1}}$, we can say that the age of the universe (${ t_{0}}$) in fourth order gravity is $0.92348\ H_{0}^{-1}$ or 13.798 Gyrs which is same as the age of the universe obtained from fitting Planck data \cite{planck} with $\Lambda$CDM model. Also, if we consider that the value of the density is given by $\rho = \rho_{d}$, a similar analysis shows that the value of $\tau_{0} = 0.9166$ and the value of ${ H_{0} =  65.0\ {\rm km \ s^{-1}Mpc^{-1}}}$, which gives the age of the universe as $0.9166\ H_{0}^{-1}$ or 13.798 Gyrs.

All the figures plotted in this Section and in the subsequent Sections, except Figs. \ref{fig:densityplotlambda} $\&$ \ref{fig:densitylambda}, are for the case when $\rho = \rho_{b}$. The results do not change significantly if we use $\rho = \rho_{d}$ instead. Hence, we have shown the results with one of the density values.
  
\begin{figure}[ht!]
\centering
\includegraphics[scale=0.7]{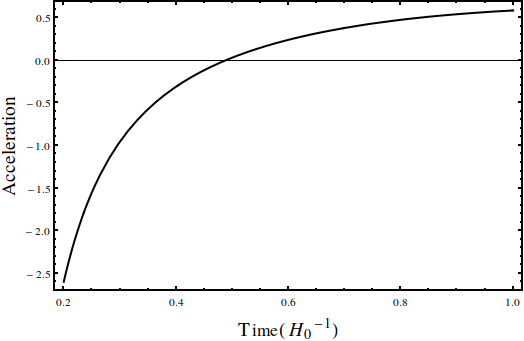}
\caption{\small Variation of acceleration of the expansion rate of the Universe with time (from $ 0.2\ { H_{0}^{-1}} $ to $ 1.0\ { H_{0}^{-1}} $) for Fourth Order Gravity.  }
\label{fig:accelesm}
\end{figure}

\begin{figure}[h!]
    \centering
        \includegraphics[width=0.7\textwidth]{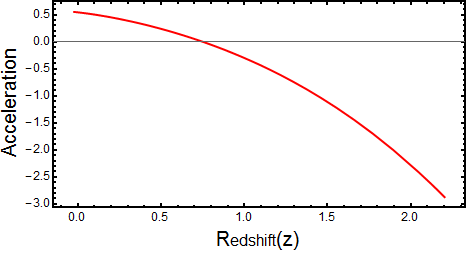}
   \caption{\small Variation of acceleration of the expansion rate of the Universe with redshift (from 0 to 2.4) for Fourth Order Gravity.}
    \label{fig:accelz}
\end{figure}
 
Fig. \ref{fig:accelesm} shows the plot of the variation of the acceleration of the universe with time for $\rho = \rho_{b}$. This plot has been obtained by numerically solving Eqn. (\ref{9}) with the initial conditions given by Eqn. (\ref{11}). Fig. \ref{fig:accelz} shows the plot of the acceleration of the universe with redshift z.
From Fig. \ref{fig:accelesm} and Fig. \ref{fig:accelz}, we see that the transition from a decelerating phase to an accelerating phase of the universe occurs at an epoch of about $0.488\ { H_{0}^{-1}}$.   
\begin{figure}[h!]
    \centering
        \includegraphics[width=0.7\textwidth]{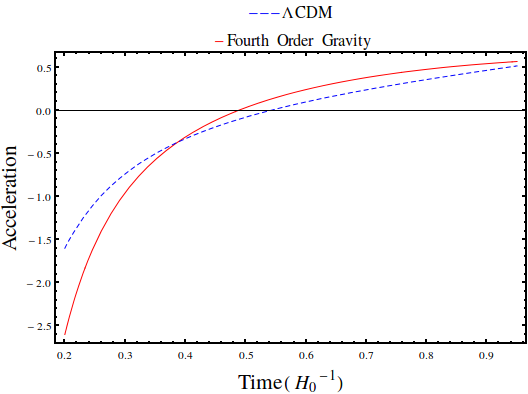}
   \caption{\small Variation of acceleration of the expansion rate of the Universe with time (from $0.2\ { H_{0}^{-1}}$ to $0.95\ { H_{0}^{-1}}$) for both Fourth Order Gravity and $\Lambda$CDM models. The blue dashed curve corresponds to $\Lambda$CDM and the red curve is for Fourth Order Gravity.}
    \label{fig:combineaccs}
\end{figure}
\begin{figure}[h!]
    \centering
        \includegraphics[width=0.7\textwidth]{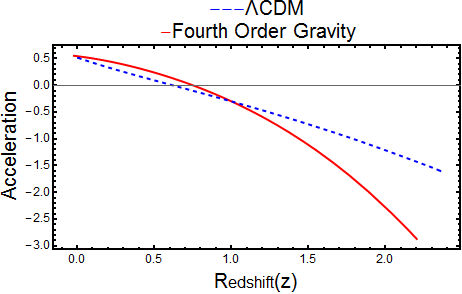}
   \caption{\small Variation of acceleration of the expansion rate of the Universe with redshift (from 0 to 2.4) for both Fourth Order Gravity and $\Lambda$CDM models. The blue dashed curve corresponds to $\Lambda$CDM and the red curve is for Fourth Order Gravity.}
    \label{fig:combineaccz}
\end{figure}
\begin{figure}[ht!]
\centering
\includegraphics[scale=0.8]{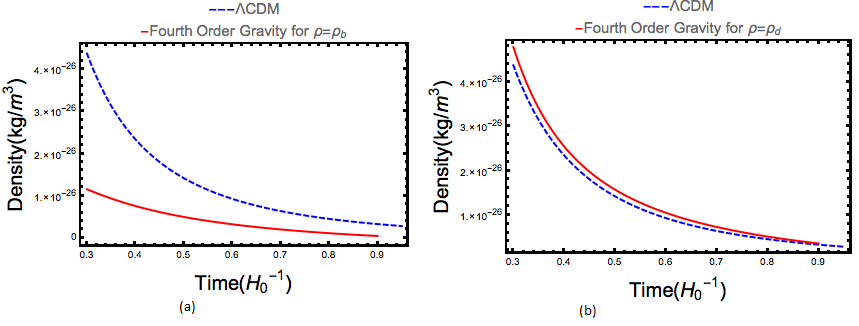}
\caption{\label{fig:densityplotlambda}\small  a) Variation of density with time (from $0.2\ { H_{0}^{-1}}$ to $0.9\ { H_{0}^{-1}}$) for Fourth Order Gravity for $\rho = \rho_{b}$ and $\Lambda$CDM model. The blue dashed curve is for $\Lambda$CDM and the red curve is for Fourth Order Gravity. b) Variation of density with time (from $0.2\ { H_{0}^{-1}}$ to $0.9\ { H_{0}^{-1}}$) for Fourth Order Gravity for $\rho = \rho_{d}$ and $\Lambda$CDM model. The blue dashed curve is for $\Lambda$CDM and the red curve is for Fourth Order Gravity.  }
\end{figure}

In Fig. \ref{fig:combineaccs} and Fig. \ref{fig:combineaccz}, we have plotted the variation of acceleration with time and redshift (from past to present time) for both $\Lambda$CDM and fourth order gravity. We find that while in $\Lambda$CDM, the scale factor is entering the accelerating phase for the first time at an epoch of around $\approx 0.54{ H_{0}^{-1}}$ (redshift of 0.61), in fourth order gravity the universe is entering an accelerating phase at an epoch of around $\approx 0.488\ { H_{0}^{-1}}$ (redshift of 0.64).

From Figs. \ref{fig:combineaccs} and \ref{fig:combineaccz}, we see that the behaviour of the acceleration of the universe from past to present epoch is nearly same for both fourth order gravity and $\Lambda$CDM model. 

 Fig. \ref{fig:densityplotlambda}(a) and Fig. \ref{fig:densityplotlambda}(b) compare the density evolution for $\Lambda$CDM and fourth order gravity from past to present epoch for both the cases i.e for $\rho = \rho_{b}$ and $\rho = \rho_{d}$. From these figures we find that the density evolution follows that of the $\Lambda$CDM model from past to present epoch. 
 
 However if we extend the density plot to future epochs, we find that the matter density acquires negative values between $1.2\ { H_{0}^{-1}}$ to $3.6\ { H_{0}^{-1}}$ and then again becomes positive. The presence of negative matter density shows that this phenomenological model is not valid for future epochs. However, since
negative density is in the future we can say this fourth order gravity model is valid up to present epoch so that acceleration is achieved. 
  
If we solve the modified Friedmann equations relaxing the constraint of pressureless matter-dominated epoch, i.e. solve them simultaneously for radiation dominated epoch in future, we find that the radiation density dominates over matter density in future in fourth order gravity model, but here also the radiation density becomes negative in future. Thus, we can conclude from the above observation that the fourth order gravity model cannot be used for future radiation dominated phase and is valid only during matter dominated phase. It remains to be understood if these fourth order corrections represent an effective modification which is a consequence of structure formation, and whether the model needs to be modified further as structures evolve in the future.

\subsection{Power law solution for the scale factor}
We can divide the evolution of scale factor with time into two phases (i) for ${ t}\ll{ H_{0}^{-1}}$ (ii) for ${ t}\geq{ H_{0}^{-1}}$.
(i) For ${ t}\ll{ H_{0}^{-1}}$, the modifying gravity terms (i.e. ${ F_{1}}$ and ${ F_{2}}$) in Eqn. (\ref{4}) and Eqn. (\ref{5}) can be neglected and these equations reduce to the standard FLRW equations. We know that for matter dominated era in FLRW Universe, the scale factor varies with time as ${ t^{2/3}}$. Hence we should expect that, for small times, the modified 
FLRW metric for fourth order gravity reduces to the standard FLRW metric with the scale factor obeying ${ t^{2/3}}$ solution.
(ii) For ${ t}\geq{ H_{0}^{-1}}$, we can assume a power law solution of the scale factor of the form
\begin{equation}
{\rm a(t) = B(t-L_{U}/c)^{n}}
\label{power}
\end{equation}
where $B$ is the proportionality constant and ${ L_{U} = c/H_{0}}$. 

Substituting Eqn. (\ref{power}) into Eqn. (\ref{8}), and neglecting the terms coming from standard FLRW equation (since here we are in the epoch where the dominating effect is due to the fourth order gravity terms), we get an equation, after some simplifications, of the form
\begin{equation}
4n^{2}-8n+3 = 0
\end{equation}
Solving the above equation for $n$, we get $n = 3/2, 1/2$ as the two solutions. But if we plot these two solutions up to the present epoch, we find that $n=3/2$ is the dominating solution. Hence we will take only the dominating solution and set $n=3/2$.
Substituting this solution in Eqn. (\ref{power}), we get
\begin{equation}
{\rm a(t) = B(t-L_{U}/c)^{3/2}}
\label{power1}
\end{equation}
\begin{figure}[h!]
    \centering
        \includegraphics[width=1.05\textwidth]{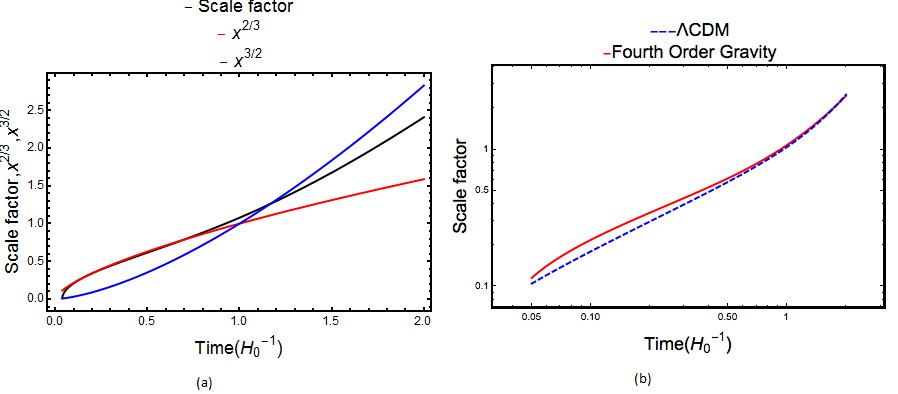}
   \caption{\small a) Variation of scale factor, ${ t^{2/3}}$ and ${ t^{3/2}}$ with time  (from $0.05\ { H_{0}^{-1}}$ to $2.0 \ { H_{0}^{-1}}$). The black curve is for the scale factor, the red line is for the ${ t^{2/3}}$ curve and the blue curve is for ${ t^{3/2}}$. b) Log-Log plot for variation of scale factor with time (from $0.05\ { H_{0}^{-1}}$ to $2.0 \ { H_{0}^{-1}}$) for both Fourth Order Gravity model and $\Lambda$ CDM model. The red curve is for Fourth Order Gravity and the blue dashed curve is for $\Lambda$ CDM model.  }
    \label{fig:powerscale}
\end{figure}
In order to get the proportionality constant, we set the left hand side of Eqn. (\ref{power1}) as the scale factor today which is equal to one and set $T$ as the current age of the universe in fourth order gravity
to get ${ B = 1/(T-L_{U}/c)^{3/2}}$.
Substituting the above expression for $B$ into Eqn. (\ref{power1}), we get
\begin{equation}
{\rm a(t) = \left(\frac{t-L_{U}/c}{T-L_{U}/c}\right)^{3/2}}
\end{equation}
 Thus the scale factor now obeys ${ t^{3/2}}$ solution.
In Fig. \ref{fig:powerscale}(a), we have plotted the variation of the scale factor with time and the curves for ${ t^{2/3}}$ and ${ t^{3/2}}$ together for comparison. It can be clearly seen from the figure that the evolution of the scale factor follows ${ t^{2/3}}$ law for ${ t}\ll{ H_{0}^{-1}}$ and approaches ${ t^{3/2}}$ law for ${ t}\geq{ H_{0}^{-1}}$. 
\begin{figure}[h!]
    \centering
        \includegraphics[width=1.1\textwidth]{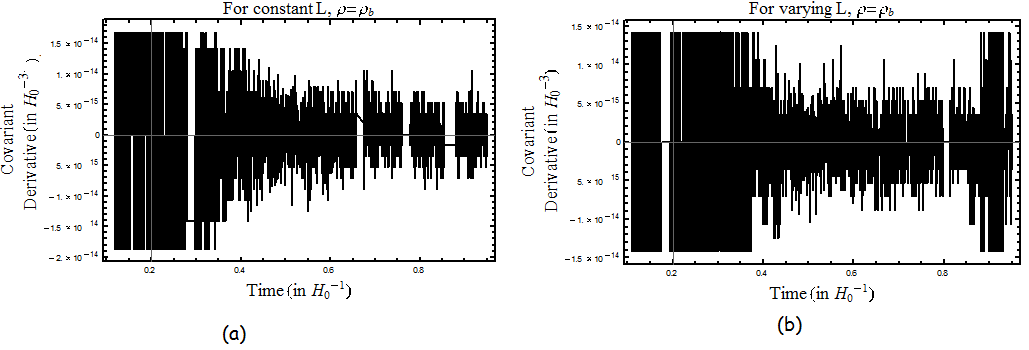}
   \caption{\small Variation with respect to time, of the expression for the covariant derivative given by Eqn. (\ref{covexp}). y-axis is plotted in units of $H_{0}^{-3}$ and x-axis  in units of $H_{0}^{-1}$. a) Plot for Fourth Order Gravity with constant length parameter $L$ and $\rho=\rho_{b}$. b) Plot for Fourth Order Gravity with varying length parameter $L(t)$ and $\rho=\rho_{b}$. Within the limits of numerical accuracy, the expression (\ref{covexp}) is zero at all times.}
    \label{fig:covariant}
\end{figure}
\begin{figure}[h!]
\centering
\includegraphics[scale=0.75]{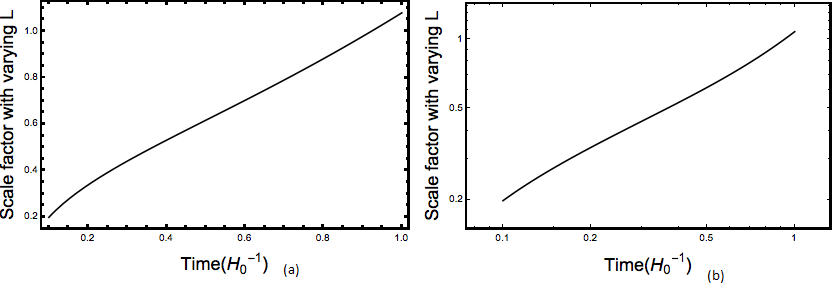}
\caption{\small a) Variation of scale factor (a(t)) with time (from $0.2\ { H_{0}^{-1}}$ to $1.0\ { H_{0}^{-1}}$) for Fourth Order Gravity with varying length parameter (L). b) Log-Log plot of scale factor (a(t)) with time (from $0.1\ { H_{0}^{-1}}$ to $1.0\ { H_{0}^{-1}}$) for Fourth Order Gravity with varying length parameter (L). }
\label{fig:newscale}
\end{figure}
\begin{figure}[h!]
\centering
\includegraphics[scale=0.7]{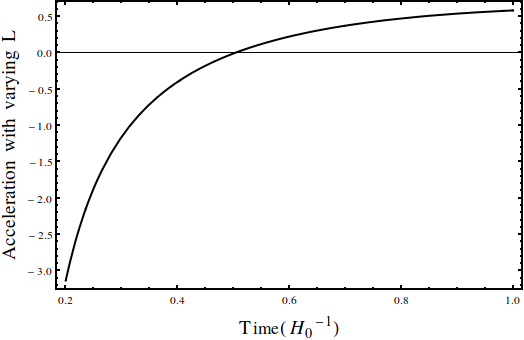}
\caption{\small Variation of acceleration of the expansion rate of the Universe with time (from $ 0.2\ { H_{0}^{-1}} $ to $ 1.0\ { H_{0}^{-1}} $) for Fourth Order Gravity with varying length parameter (L).}
\label{fig:newaccel}
\end{figure}
\begin{figure}[h!]
\centering
\includegraphics[scale=0.6]{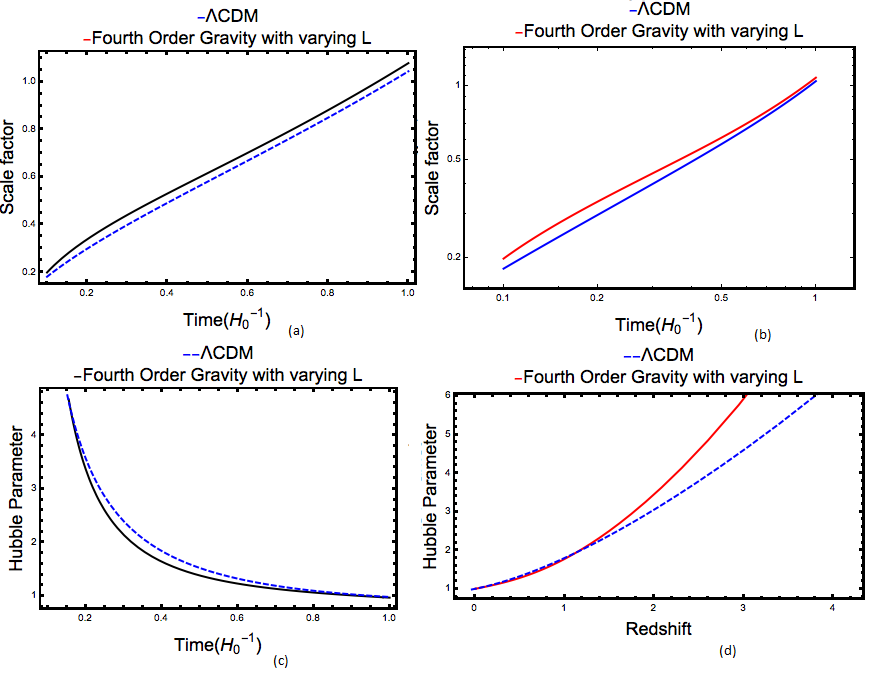}
\caption{\label{fig:fig6}\small a) Variation of scale factor (a(t)) with time (from $0.1\ { H_{0}^{-1}}$ to $0.95\ { H_{0}^{-1}}$) for Fourth Order Gravity with varying length parameter (L) and $\Lambda$CDM. The blue-dashed curve is for $\Lambda$CDM and the black curve is for Fourth Order Gravity. b)Log-Log plot of scale factor (a(t)) with time (from $0.1\ { H_{0}^{-1}}$ to $0.95\ { H_{0}^{-1}}$) for Fourth Order Gravity with varying length parameter (L) and $\Lambda$CDM. The blue curve is for $\Lambda$CDM and the red curve is for Fourth Order Gravity. c) Variation of the Hubble parameter with time (from $0.1\ { H_{0}^{-1}}$ to $0.95\ { H_{0}^{-1}}$) for Fourth Order Gravity with varying length parameter (L) and $\Lambda$CDM. The blue-dashed curve is for $\Lambda$CDM and the black curve is for Fourth Order Gravity. d) Variation of the Hubble parameter with redshift (from 0 to 4.6) for Fourth Order Gravity with varying length parameter (L) and $\Lambda$CDM. The blue-dashed curve is for $\Lambda$CDM and the red curve is for Fourth Order Gravity. }
\end{figure}
\begin{figure}[h!]
\centering
\includegraphics[scale=0.75]{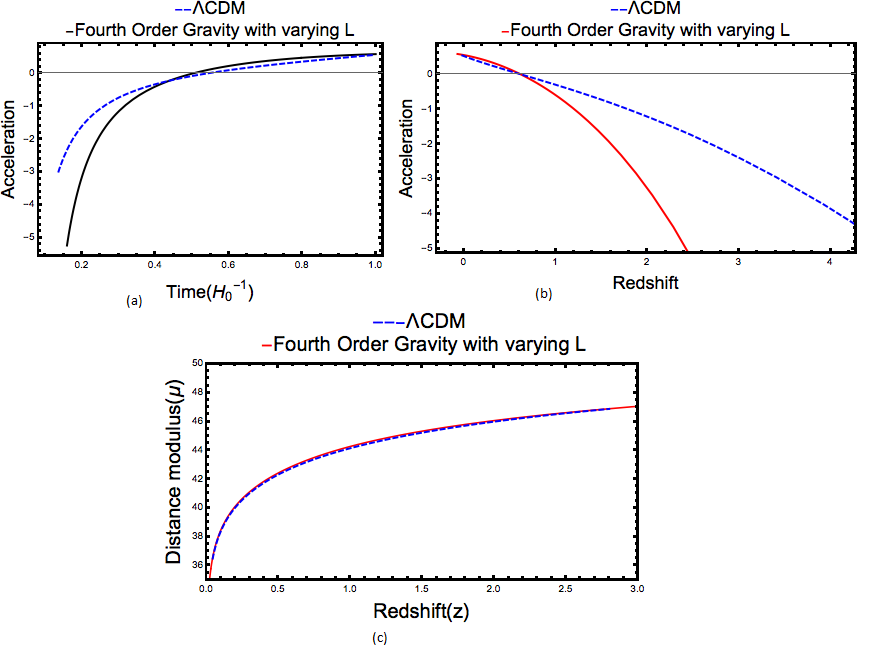}
\caption{\label{fig:fig7}\small a) Variation of acceleration of the expansion of the Universe with time (from $0.1 \ { H_{0}^{-1}}$ to $0.95 \ { H_{0}^{-1}}$) for Fourth Order Gravity with varying length parameter (L) and $\Lambda$CDM. The blue-dashed curve is for $\Lambda$CDM and the black curve is for Fourth Order Gravity. b) Variation of acceleration of the expansion of the Universe with redshift (from 0 to 3.0) for Fourth Order Gravity with varying length parameter (L) and $\Lambda$CDM. The blue-dashed curve is for $\Lambda$CDM and the red curve is for Fourth Order Gravity. c) Hubble diagram for Fourth Order Gravity with varying length parameter (L)and $\Lambda$CDM.  The blue-dashed curve is for $\Lambda$CDM and the red curve is for Fourth Order Gravity.  }
\end{figure}
\begin{figure}[ht!]
\centering
\includegraphics[scale=1.13]{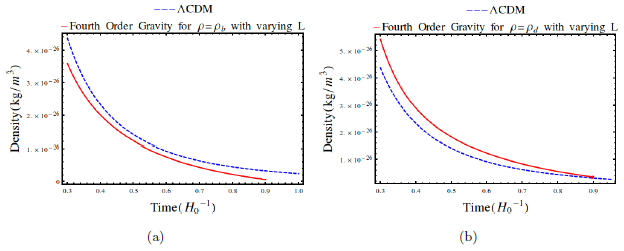}
\caption{\label{fig:densitylambda}\small  a) Variation of density with time (from $0.3\ { H_{0}^{-1}}$ to $1.0\ { H_{0}^{-1}}$) for Fourth Order Gravity with varying L for $\rho = \rho_{b}$ and $\Lambda$CDM model. The blue dashed curve is for $\Lambda$CDM and the red curve is for Fourth Order Gravity.  b) Variation of density with time (from $0.3\ { H_{0}^{-1}}$ to $0.95\ { H_{0}^{-1}}$) for Fourth Order Gravity with varying L for $\rho = \rho_{d}$ and $\Lambda$CDM model. The blue dashed curve is for $\Lambda$CDM and the red curve is for Fourth Order Gravity.  }
\end{figure}
In Fig. \ref{fig:powerscale}(b), we have plotted the variation of scale factor from past to near future i.e from $t<H_{0}^{-1}$ to $t>H_{0}^{-1}$ for both fourth order gravity model and $\Lambda$CDM model. From the figure we can see that the variation of scale factor for both the model is nearly the same.

\section{Effect of considering the variation of the length parameter (L) with time}
In Section II, we had set the length parameter ${ L = c/H_{0}}$. It is more realistic to take into account the variation of the Hubble parameter, and assume  ${ L(t) = c/H(t)}$ where ${ H(t) = \dot{a}/a}$. 
Using this condition, we once again follow the same procedure as in Section II and get the new modified Friedmann equations as,
\beq
\frac{\dot{{\rm a}}^{2}}{{\rm a}^{2}}+\frac{1}{{\rm H}^{2}}{\rm F}_{1}({\rm a},\dot{{\rm a}},\ddot{{\rm a}},\dddot{{\rm a}}) = {\rm \frac{8\pi G}{3c^{2}}}\rho
\label{80}
\eeq
\beq
\frac{2\ddot{{\rm a}}}{{\rm a}}+\frac{\dot{{\rm a}}^{2}}{{\rm a}^{2}}+\frac{1}{{\rm H}^{2}}{\rm F}_{2}({\rm a},\dot{{\rm a}},\ddot{{\rm a}},\dddot{{\rm a}},\ddddot{{\rm a}}) = -{\rm\frac{8\pi G}{c^{2}}p}
\label{90}
\eeq 
  where the explicit forms of ${ F_{1}}$ and ${ F_{2}}$ are
  \beq
{\rm F}_{1} = \frac{{\rm a}^{2}\dot{{\rm a}}\dddot{{\rm a}}+{\rm a}\dot{{\rm a}}^{2}\ddot{{\rm a}}-2\dot{{\rm a}}^{4}}{{\rm a}^{4}}
\label{6}
\eeq
\beq
{\rm F}_{2} = \frac{{\rm a}^{3}\ddddot{{\rm a}}+2{\rm a}^{2}\dot{{\rm a}}\dddot{{\rm a}}+{\rm a}^{2}\ddot{{\rm a}}^{2}-6{\rm a}\dot{{\rm a}}^{2}\ddot{{\rm a}}+2\dot{{\rm a}}^{4}}{{\rm a}^{4}}
\label{7}
\eeq
 
 Setting ${ H = \dot{a}/a}$ and solving Eqn. (\ref{90}) for a non-relativistic (i.e. pressureless) and matter-dominated Universe, gives 
\begin{equation}
2{\rm a^{2}\dot{a}\dddot{a}}+{\rm a^{2}\ddot{a}^{2}}+{\rm a^{3}\ddddot{a}}+3{\rm \dot{a}^{4}}-4{\rm a\dot{a}^{2}\ddot{a}} = 0
\label{91}
\end{equation}
Once again applying the coordinate transformation given by Eqn. (\ref{8.1}) to Eqn. (\ref{91}), we get the modified Friedmann equation in the new time coordinate as
\begin{equation}
2{\rm a^{2}a'a'''}+{\rm a^{2}a''^{2}}+{\rm a^{3}a''''}+3{\rm a'^{4}}-4{\rm aa'^{2}a''} = 0
\end{equation}
where $'$ denotes derivative with respect to $\tau$.

Before proceeding further we must address an important issue: the covariance of the right hand side of the field equations  (\ref{2}) when we introduce a time-dependent length scale $L(t) = c/H(t)$. Such an $L(t)$ explicitly depends on the Robertson-Walker time coordinate $t$, thus apparently breaking covariance. However the correct way to think of such an $L(t)$ is in terms of the expansion scalar 
$\Theta$ for a congruence of spherically expanding time-like geodesics. For a Robertson-Walker spacetime, the scalar takes the value $\Theta = 3H(t)$, and hence $L(t)=3c/\Theta$. Therefore the covariant expression for the field equations (\ref{2}) is
\beq
{\rm R}_{\mu\nu}-\frac{1}{2}{\rm g}_{\mu\nu}{\rm R} = {\rm \frac{8\pi G}{c^4}T}_{\mu\nu}+ \left({\frac{3c}{\Theta}}\right)^{2}{\rm R}_{\mu\alpha\nu\beta}^{;\alpha\beta}
\label{cov}
\eeq
For the special case considered earlier, where $L$ was a constant, we interpret the expansion scalar as having been set to its value at the present epoch.

Next, we must ask if the right hand side of Eqn. (\ref{cov}) is covariantly conserved, as it must be, since the left hand side is conserved, by virtue of the Bianchi identities. The right hand side is a symmetric second rank tensor, which we denote as $\Psi_{\mu\nu}$:
\beq
\Psi_{\mu\nu} \equiv   {\rm \frac{8\pi G}{c^4}T}_{\mu\nu}+ \left({\frac{3c}{\Theta}}\right)^{2}{\rm R}_{\mu\alpha\nu\beta}^{;\alpha\beta}
\label{rhs}
\eeq
We need to show that $\Psi_{\mu\nu}^{\ \ ;\nu}=0$. We expect this to be true so long as all the field equations are solved simultaneously. In particular, this will be true in the Robertson-Walker case if the two Friedmann equations are solved simultaneously for the scale factor and matter density (we are considering the pressureless case). We now demonstrate this explicitly for the Friedmann equations. It is straightforward to check that the only non-trivial component of $\Psi_{\mu\nu}^{\ \ ;\nu}$ is $\Psi_{t\nu}^{\ \ ;\nu}$. The other components, $\Psi_{\i\nu}^{\ \ ;\nu}$, where $i$ is a spatial index, can be shown to vanish identically, for constant $L$ as well as time-dependent $L$.  For time-dependent $L$, the non-trivial component is given by the expression
\beq
\Psi_{t\nu}^{\ \ ;\nu} = \frac{8\pi G}{3} \left( \dot\rho + 3 \frac{\dot a}{a}\rho \right)
-\frac{d\ }{dt} \left(\frac{F_1}{H^2}\right) - 3 \frac{\dot a}{a} \frac{F_1}{H^2} + \frac{\dot a}{a}\frac{F_2}{H^2}
\label{covexp}
\eeq 
In this equation, we substitute for the density and its time derivative from Eqn. (\ref{80}), in terms of $F_1$ and $\dot{F_1}$. We get that
\beq
  \Psi_{t\nu}^{\ \ ;\nu} = \frac{\dot a^3}{a^3} + \frac{2\dot a \ddot a}{a^2} + \frac{\dot a}{a} \frac{F_2}{H^2}
  \eeq
We then substitute for the $\ddot a$ term from Eqn. (\ref{90}) (after setting $p=0$) and then the above covariant derivative identically vanishes. A similar proof holds  for the case of constant $L$ - we only have to replace $H$ by $H_0$ in (\ref{covexp}) and repeat the  same argument using the field equations (\ref{4}) and (\ref{5}) for constant $L$. Thus we have shown that the right hand side of the field equations is covariant, and covariantly conserved, when the scale factor and density simultaneously satisfy the two Friedmann equations. For additional confirmation, we have plotted the expression (\ref{covexp}) as a function of time, in Fig. \ref{fig:covariant}, for the solution that we have worked out for the scale factor. We have used $\rho=\rho_b$; similar results hold for $\rho=\rho_d$. It is evident from the figure that, within the limits of numerical accuracy, this expression is zero at all times.

We now return to the analysis of the time dependent $L(t)$ case. Following the procedure of the previous sections and using the same initial conditions, we repeat the above calculations and find that most of the results of Section III and IV still hold true i.e. fourth order gravity still gives a good fit to the luminosity distance curve. The success of this version of the model, which employs a varying $H(t)$, suggests that this model provides a way to address the cosmic coincidence problem: there is nothing special about today's epoch in this model.

In Figs. \ref{fig:newscale}-\ref{fig:newaccel}, we have shown the evolution of the scale factor and the  acceleration of the scale factor with time. In this Section also, all the figures are for the case when $\rho = \rho_{b}$.

 From  Fig. \ref{fig:newaccel}, we find that the universe enters an accelerating phase just as it was doing earlier when we had considered a constant Hubble parameter.

 Fig. \ref{fig:fig7} and Fig. \ref{fig:fig6} show the plots comparing the past evolution of the scale factor, Hubble parameter, acceleration and the distance modulus in fourth order gravity with varying length parameter and $\Lambda$CDM model. From these plots, we can infer that the past evolution of the universe in fourth order gravity agrees well with that of $\Lambda$CDM model, even with a varying length parameter.

 Fig. \ref{fig:densitylambda}(a) and Fig. \ref{fig:densitylambda}(b)  compare the density evolution for $\Lambda$CDM and fourth order gravity with varying $L$ from past to present epoch  for both the cases i.e. when $\rho = \rho_{b}$ and $\rho = \rho_{d}$ in fourth order gravity. From these figures we find that the density evolution follows that of the $\Lambda$CDM model from past to present epoch just as it was doing for constant $L$ case.
 
 The conclusions regarding the negative density epoch and radiation domination in future in fourth order gravity model that we had got earlier for the constant $L$ case, remains valid even if we vary the length parameter.

\section{Comparing correction terms in Fourth Order Gravity and $\Lambda$CDM}
It can be shown by explicit computation that the correction term $R^{;\alpha\beta}_{\mu\alpha\nu\beta}$
is diagonal, suggesting that it could be proportional to the metric, and hence effectively behaves like the cosmological constant term in the $\Lambda$CDM model. To verify this, we recall that the
Friedmann equations in the $\Lambda$CDM model in our new coordinate frame i.e. $\tau = tH_0$, are given by
\begin{eqnarray}
{\rm \frac{a'}{a}^{2} - \frac{\Lambda c^2}{3H_{0}^{2}}} = {\rm \frac{8\pi G}{3H_{0}^{2}}}\rho \\
{\rm \frac{2a''}{a} + \frac{a'}{a}^{2} - 3\frac{\Lambda c^2}{3H_{0}^{2}}} = -{\rm\frac{8\pi G}{c^{2}H_{0}^{2}}p}
\end{eqnarray}
Therefore expressing $\Lambda$ in terms of the cosmological constant density parameter $\Omega_{\Lambda}$, the correction terms in $\Lambda$CDM model for the Friedmann equations are -$\Omega_{\Lambda}$ and -$3\Omega_{\Lambda}$ respectively. Using the values of $\Omega_{\Lambda}$ from Planck data we get the magnitude of the corresponding correction terms as -0.68 and -2.4 respectively.

In the fourth order gravity model, the Friedmann equations in terms of $\tau$ are given by Eqn. (\ref{4new}) and Eqn. (\ref{5new}) and the corresponding correction terms are given by Eqn. (\ref{6new}) and Eqn. (\ref{7new}) respectively.
\begin{figure}[ht!]
\centering
\includegraphics[scale=1.0]{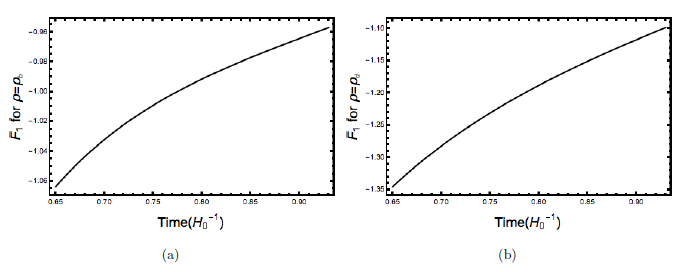}
\caption{\label{fig:f1}\small  a) Variation of $F_{1}$ with time (from $0.65\ { H_{0}^{-1}}$ to $0.93\ { H_{0}^{-1}}$) for Fourth Order Gravity with constant L for $\rho = \rho_{b}$.  b) Variation of $F_{1}$ with time (from $0.65\ { H_{0}^{-1}}$ to $0.93\ { H_{0}^{-1}}$) for Fourth Order Gravity with constant L for $\rho = \rho_{d}$.  }
\end{figure}

\begin{figure}[ht!]
\centering
\includegraphics[scale=0.95]{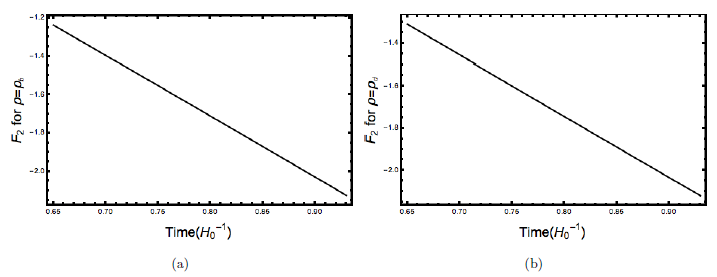}
\caption{\label{fig:f2}\small  a) Variation of $F_{2}$ with time (from $0.65\ { H_{0}^{-1}}$ to $0.93\ { H_{0}^{-1}}$) for Fourth Order Gravity with constant L for $\rho = \rho_{b}$.  b) Variation of $F_{2}$ with time (from $0.65\ { H_{0}^{-1}}$ to $0.93\ { H_{0}^{-1}}$) for Fourth Order Gravity with constant L for $\rho = \rho_{d}$. }
\end{figure}

Figs. \ref{fig:f1}(a) and \ref{fig:f1}(b) show the evolution of $F_1$ from near past to present epoch for fourth order gravity with constant L for $\rho = \rho_b$ and $\rho = \rho_d$ respectively. Comparing it with the corresponding correction term in $\Lambda$CDM which is a constant and is given by -0.68, we see that the correction term in fourth order gravity model is nearly constant with its magnitude varying from -0.96 to -1.064 for $\rho = \rho_{b}$ and from -1.101 to -1.35 for $\rho = \rho_{d}$.  

Figs. \ref{fig:f2}(a) and \ref{fig:f2}(b) show the evolution of $F_2$ from near past to present epoch for fourth order gravity with constant L for $\rho = \rho_b$ and $\rho = \rho_d$ respectively. Comparing it with the corresponding correction term in $\Lambda$CDM which is a constant and given by -2.4, we see that the correction term in fourth order gravity model is nearly constant with its magnitude varying from -1.24 to -2.13 for $\rho = \rho_{b}$ and from -1.3 to -2.12 for $\rho = \rho_{d}$.
 
 The results remain same if we repeat the above analysis for varying length parameter in fourth order gravity. These results suggest that the correction terms in the fourth order model effectively behave nearly, but not exactly, like the cosmological constant. The difference is brought out by studying the equation of state in our model, as done in the next section. 

\section{Effective equation of state in Fourth Order Gravity}
With a suitable rewriting of the correction terms in fourth order gravity, we may treat them as an effective dark energy and work out the corresponding equation of state.
We can rewrite Eqn. (\ref{2}) (with c=1) as 
\begin{equation}
{\rm R}_{\mu\nu}-\frac{1}{2}{\rm g}_{\mu\nu}{\rm R} = 8\pi {\rm G}\left[{\rm T}_{\mu\nu}\mid_{{\rm matter}} + {\rm T}_{\mu\nu}\mid_{{\rm DE}}\right]
\label{DE}
\end{equation}
where $  T_{\mu\nu}\mid_{{ DE}} = \frac{{ L}^{2}}{8\pi G}{ R}_{\mu\alpha\nu\beta}^{;\alpha\beta}$.

Similarly we can also write the Friedmann equations  given by Eqns. (\ref{4}) and (\ref{5}) for pressureless matter-dominated universe as 
\begin{equation}
\frac{\dot{{\rm a}}^{2}}{{\rm a}^{2}} = {\rm \frac{8\pi G}{3}}\rho_{{\rm matter}} +    {\rm \frac{8\pi G}{3}}\rho_{{\rm DE}}
\label{DE1}
\end{equation}
\begin{equation}
\frac{2\ddot{{\rm a}}}{{\rm a}}+\frac{\dot{{\rm a}}^{2}}{{\rm a}^{2}} = -{\rm 8\pi G}{\rm p}_{{\rm DE}}
\label{DE2}
\end{equation}
where 
\begin{eqnarray}
\rho_{{\rm DE}} &=&  -{\rm \frac{3}{8\pi G}} \frac{1}{{\rm H}_{0}^{2}}{\rm F}_{1}({\rm a},\dot{{\rm a}},\ddot{{\rm a}},\dddot{{\rm a}}) \\
{\rm p}_{{\rm DE}} &=& {\rm\frac{1}{8\pi G}}\frac{1}{{\rm H}_{0}^{2}}{\rm F}_{2}({\rm a},\dot{{\rm a}},\ddot{{\rm a}},\dddot{{\rm a}},\ddddot{{\rm a}})
\end{eqnarray}
where the expressions for $F_{1}$ and $F_{2}$ are given by Eqns. (\ref{6}) and (\ref{7}) respectively.

Therefore the effective equation of state for dark energy in fourth order gravity model is given by
\begin{equation}
w_{{\rm DE}} = {\rm \frac{p_{DE}}{\rho_{DE}}} = - {\rm \frac{F_{2}}{3F_{1}}}
\label{DE3}
\end{equation}
\begin{figure}[ht]
\centering
\includegraphics[scale=0.7]{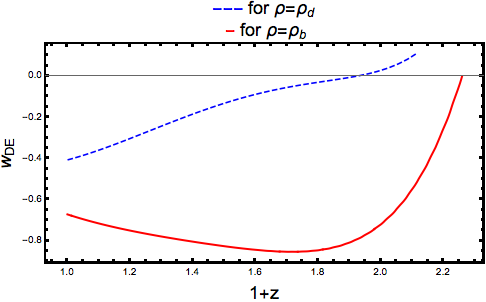}
\caption{\small  Equation of state parameter for Fourth Order Gravity.  
}
\label{fig:wde}
\end{figure}
We can see  from Eqn. (\ref{DE3}), that the equation of state does not have any explicit dependence on the length parameter L, it  depends on it implicitly through the solution of the scale factor. Rewriting Eqn. (\ref{DE3}) in terms of $\tau=tH_{0}$, and using the numerical solution of $a(\tau)$, we have plotted in Fig. \ref{fig:wde}, the variation of the equation of state $(w_{DE})$ with redshift (1+z) for both the cases i.e. when $\rho = \rho_{b}$ and $\rho = \rho_{d}$. The equation of state parameter for $\Lambda$CDM model is constant from past to present epoch and the value of $w$ is -1. But as we can see the fourth order gravity model predicts an evolving equation of state whose value varies from -0.85 to -0.5 from near past to present redshift for $\rho = \rho_{b}$ and from -0.4 to +0.1 for $\rho = \rho_{d}$. Therefore we can conclude that in fourth order  gravity, if we consider that $\rho$ is only made up of baryons, then $-1<w_{DE}<0$ but if we also include dark matter then $w_{DE} > 0$ is also possible. The present value of $w_{DE}$ for $\rho = \rho_{b}$ is -0.68 and for $\rho = \rho_{d}$ is -0.4.

We can also parametrize $w_{DE}$, given by Eqn. (\ref{DE3}), with respect to redshift (z) using the following Taylor series expansion:

For $\rho = \rho_{b}$
\begin{equation}
w_{DE} = w_{0} + w_{1}{\rm z} +  w_{2}{\rm z}^{2} + w_{3}{\rm z}^{3} + w_{4}{\rm z}^{4} .
\end{equation}
where $w_{0}\approx -0.68, w_{1} = \frac{dw_{DE}}{dz}\mid_{z=0},  w_{2} = \frac{d^{2}w_{DE}}{dz^{2}}\mid_{z=0},  w_{3} = \frac{d^{3}w_{DE}}{dz^{3}}\mid_{z=0}   \ {\rm and}\  w_{4} = \frac{d^{4}w_{DE}}{dz^{4}}\mid_{z=0} $.

For $\rho = \rho_{d}$
\begin{equation}
w_{DE} = w_{0} + w_{1}{\rm z} +  w_{2}{\rm z}^{2} .
\end{equation}
where $w_{0}\approx -0.4, w_{1} = \frac{dw_{DE}}{dz}\mid_{z=0} \ {\rm and}\  w_{2} = \frac{d^{2}w_{DE}}{dz^{2}}\mid_{z=0}$.

A more sensitive diagnosis of the present accelerating epoch could be done in terms of the state finders, $(r, s)$, first proposed in \cite{sahni1} \cite{sahni2}. Expressed in terms of the higher derivatives of the scale factor, they provide a geometric probe of the expansion dynamics of the universe. Their explicit forms in terms the scale factor and its derivatives are given by
\begin{eqnarray}
{\rm r} &=& {\rm \frac{\dddot{a}}{aH^{3}}} \\
{\rm s} &=& {\rm \frac{(r-1)}{3(q-1/2)}}
\end{eqnarray}
where $q$ is the deceleration parameter defined in Section II. 
\begin{figure}[ht!]
\centering
\includegraphics[scale=0.75]{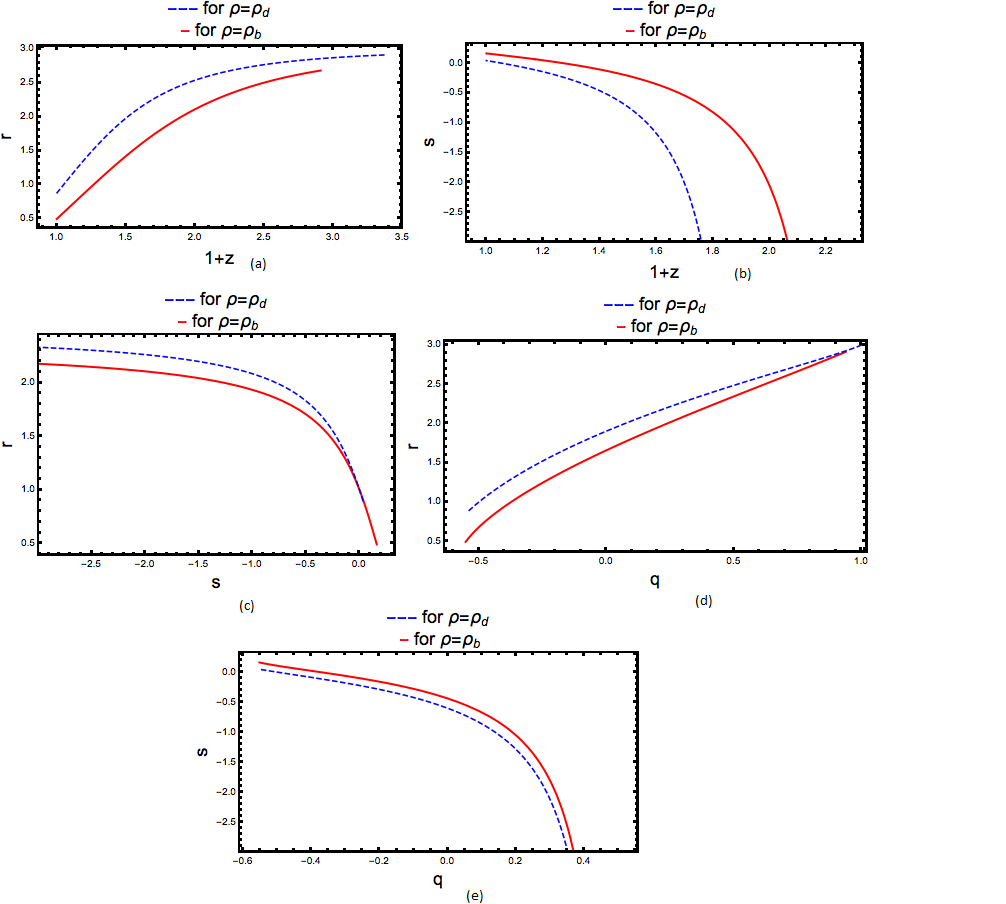}
\caption{\label{fig:r}\small  a) Variation of  state finder $r$ with redshift $(1+z)$ for $\rho = \rho_{b}$ and $\rho = \rho_{d}$ . The blue dashed curve is for $\Lambda$CDM and the red curve is for Fourth Order Gravity. b) Variation of state finder $s$ with redshift (1+z) for $\rho = \rho_{b}$  and $\rho = \rho_{d}$. c) Variation of the state finder pair $(r,s)$ in for $\rho = \rho_{b}$  and $\rho = \rho_{d}$. d) Evolution of the pair $(r,q)$ for $\rho = \rho_{b}$  and $\rho = \rho_{d}$, where $q$ is the deceleration parameter. e) Evolution of the pair $(s,q)$ for $\rho = \rho_{b}$  and $\rho = \rho_{d}$, where $q$ is the deceleration parameter. In all the above five figures, the blue dashed curve is for $\rho = \rho_{d}$ and the red curve is for $\rho = \rho_{b}$.  }
\end{figure}
Figs. \ref{fig:r}(a)-\ref{fig:r}(b) show the variations of $r$ and $s$ with respect to redshift for both the cases i.e. when $\rho = \rho_{b}$ and $\rho = \rho_{d}$. From Fig. \ref{fig:r}(a), we see that for both the cases, $r$ takes positive values only,  and it follows nearly the same evolution pattern for  both the cases. The value of $r$ at the present epoch for fourth order gravity model is 0.48 for $\rho = \rho_{b}$ and 0.86 for $\rho = \rho_{d}$. For reference, the value of $r$ in $\Lambda$CDM model is 1. We also find from Fig. \ref{fig:r}(b) that the value of $s$ is mostly negative, becoming slightly positive around the present epoch. The evolution of $s$ for both the cases is nearly the same with the present value of $s$ being $\approx 0.16$  for $\rho = \rho_{b}$ and 0.05 for $\rho = \rho_{d}$. For $\Lambda$CDM, $s$ = 0. In Figs. \ref{fig:r}(c)-\ref{fig:r}(e), we have plotted the graphs showing the trajectories in $(r-s), (r-q)$ and $(s-q)$ planes for both the cases and we see that the evolutions are almost same for both the cases i.e. when $\rho = \rho_{b}$ and $\rho = \rho_{d}$. As we can see from Fig. \ref{fig:r}(c), the point (1,0), lies on the $r$ vs $s$ curve. The present values of the state finders in fourth order gravity are (0.48, 0.16) and (0.86, 0.05) for $\rho = \rho_{b}$ and  $\rho = \rho_{d}$ respectively. Also from Figs. \ref{fig:r}(c) and \ref{fig:r}(d), we see that while in $(r-s)$ plane, the curves converge as they approach the present epoch, in $(r-q)$ plane, they converge in the past.

The above conclusions remain almost the same when we make the length parameter time dependent.

\section{Redshift drift}
In order to contrast this model with the standard model, we next compare and contrast the cosmological redshift drift amongst the two models.
Originally considered by Sandage \cite{sandage} and then by McVittie \cite{vittie}, it is a tool which is used to directly probe the expansion history of the universe without the need for any cosmological priors. The redshift drift is the temporal variation of the redshift of distant sources when the observation of the same source is done at observer's different proper times in an expanding universe. It allows one to make observations on the past light cones of an observer at different cosmological times.

We know that the general definition of the redshift of a source is given by 
\begin{equation}
{\rm z(t_{0}) =\frac{ a(t_{0})}{a(t_{e})}-1}
\label{z}
\end{equation}
where ${ t_{e}}$ is the time when the signal was emitted from the source and ${ t_{0}}$ is the time when it is observed i.e. the present time.

Since, the redshift of the source is measured on the observer's two different past light cones, after an elapsed time $\delta{ t_{0}}$, the redshift of the source is given by
\begin{equation}
{\rm z(t_{0} + \delta t_{0}) =\frac{ a(t_{0} + \delta t_{0})}{a(t_{e} + \delta t_{e})} -1}
\label{deltaz}
\end{equation}
where ${\delta t_{e}}$ is the time interval within which the source emitted another signal.
\begin{figure}[h!]
\centering
\includegraphics[scale=1.00]{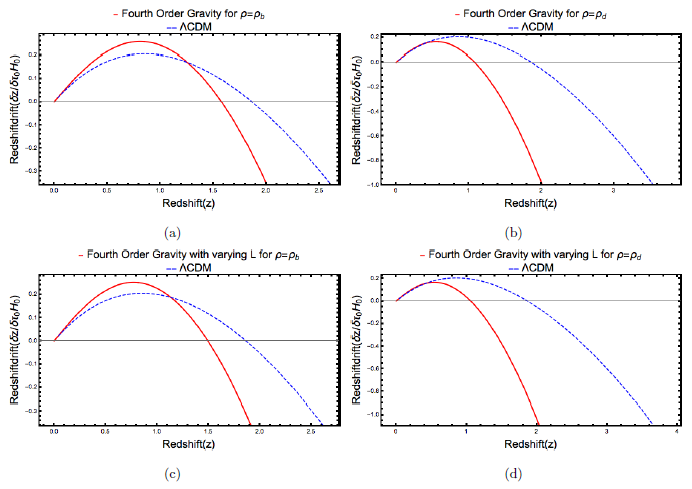}
\caption{\label{fig:drift1}\small Variation of redshift drift with redshift for $\Lambda$CDM and a) Fourth Order Gravity with $\rho=\rho_{b}$, b) Fourth Order Gravity with $\rho=\rho_{d}$, c) Fourth Order Gravity for varying L with $\rho=\rho_{b}$,  d) Fourth Order Gravity for varying L with $\rho=\rho_{d}$. In all the four plots, the blue-dashed curve is for $\Lambda$CDM and the red curve is for Fourth Order Gravity.  }
\end{figure}
Using the definition ${ \delta t_{e} = \delta t_{0}/(1 + z)}$ and subtracting Eqn. (\ref{z}) from Eqn. (\ref{deltaz}) and applying first order approximation, we get the well known McVittie equation \cite{vittie} 
\begin{equation}
{\rm \frac{\delta z}{\delta t_{0}} = (1 + z)H_{0} - H(z)}
\label{drift}
\end{equation}
where ${ H(z)  = \dot a(t_{e})/a(t_{e})}$. 

Rewriting  Eqn. (\ref{drift}) in terms of our new variable $\tau = { tH_{0}}$, we get
\begin{equation}
{\rm \frac{\delta z}{\delta \tau_{0}} = H_{0}\left[(1 + z) - \frac{a'(\tau_{e})}{a(\tau_{e})}\right]}
\end{equation}
Using the numerical solution of ${ a(\tau)}$ from Eqns. (\ref{9}) $\&$ (\ref{91})  for constant and varying length parameter respectively and for both the cases when $\rho = \rho_b$ and when $\rho = \rho_d$, we get the redshift drift in fourth order gravity model for a flat, matter-dominated universe.

In Figs. \ref{fig:drift1}(a)-\ref{fig:drift1}(d), we have compared the variation of redshift drift with redshift in fourth order gravity and $\Lambda$CDM model for both constant and varying length parameter and for both the cases when $\rho=\rho_{b}$ and $\rho=\rho_{d}$.   From the plots we can conclude that the variation of redshift drift in fourth order gravity with redshift nearly follows that in $\Lambda$CDM model. Like in $\Lambda$CDM, here also we see that for all the four cases shown, the redshift drift shows a positive variation for low redshifts because of the acceleration of the universe. But the transition from negative to positive redshift drift occurs at an earlier epoch in $\Lambda$CDM model as compared to fourth order gravity model. However, we also see that the redshift drift in fourth order gravity starts showing the negative behaviour at an earlier redshift as we increase the matter content i.e. as we include the dark matter component along with the baryons.

\section{Concluding Remarks} 
Theories of modified gravity which act as alternatives to dark energy are a useful benchmark against which the standard model can be tested, and such theories can be ruled out or confirmed by surveys such as the Dark Energy survey, which hopes to shed light on the equation of state. 
We have seen that the fourth order gravity model studied here does well in explaining cosmic acceleration, if the Hubble constant and the jerk parameter are treated as free parameters whose values are determined by the best fit to data. Of particular interest is the version of the model in which the length parameter $L$ is allowed to vary with epoch, as doing so provides an explanation for the cosmic coincidence problem. Since the evolution of the equation of state and of redshift drift is different from that in the standard model, these serve as helpful discriminators.  Work is currently in progress to see if the model will stand up to further scrutiny, as regards comparison with other data, and growth of perturbations in linear theory and matching with CMB data.

\bigskip

\noindent{\bf Acknowledgement:} We would like to thank Ken-ichi Nakao for helpful discussions.

 \newpage

\end{document}